\documentclass{article}

\usepackage{amsmath}
\usepackage{amssymb}
\usepackage{graphicx}

\begin{document}

\title{Evolutionary Stable Strategies in Games with Fuzzy Payoffs}

\author{Haozhen~Situ% <-this % stops a space
\thanks{H.Z. Situ is with the College of Mathematics and Informatics, South China Agricultural University, Guangzhou,
 510642 China. E-mail: situhaozhen@gmail.com}}% <-this % stops a space

\maketitle

\begin{abstract}
Evolutionarily stable strategy (ESS) is a key concept in evolutionary game theory.
ESS provides an evolutionary stability criterion for biological, social and economical behaviors.
In this paper, we develop a new approach to evaluate ESS in symmetric two player games with fuzzy payoffs. Particularly, every strategy is assigned a fuzzy membership that describes to what degree it is an ESS in presence of uncertainty. The fuzzy set of ESS characterize the nature of ESS. The proposed approach avoids loss of any information that happens by the defuzzification  method in games and handles uncertainty of payoffs through all steps of finding an ESS. We use the satisfaction function to compare fuzzy payoffs, and adopts the fuzzy decision rule to obtain the membership function of the fuzzy set of ESS. The theorem shows the relation between fuzzy ESS and fuzzy Nash equilibrium. The numerical results illustrate the proposed method is an appropriate generalization of ESS to fuzzy payoff games.
\end{abstract}


\section{Introduction}\label{s1}

The general problem of how to make decisions plays an important role in economics, biology, political science, computer science and etc. Generally in real life problems, there are several decision makers who affects the actions of the others. Game theory is the study of mathematical models of conflict and cooperation between rational decision-makers. A normal game consists of a set of players (decision makers), their strategies (actions), and payoffs available for all combinations of the players' strategies. Classical game theory \cite{Basar1998,Fudenberg1991} deals with rational players that are engaged in a game with other players. Each player has to decide among his strategies in order to maximize his payoff, which depends on the strategies chosen by other players who in turn try to maximize their payoffs. However, rationality is generally a strong assumption imposed to real world scenarios. As a result, evolutionary game theory \cite{Weibull1995,Hofbauer1998} (EGT) deals with a population of players, in which players are not assumed to play rationally, with rationality replaced by evolutionary stability. The players in EGT are repeatedly drawn from a random infinite population to play the game.  The aim is to study the evolution of the different strategies in the population according to a behavioral pattern.

The main concept in EGT is that of an evolutionary stable strategy (ESS) \cite{Maynard1973} describing the stable state of a population resulting from dynamics of evolution. Such a strategy is robust to evolutionary selection forces in an exact sense. Initially all individuals play the same incumbent strategy, then a small population share of individuals who play some other mutant strategy is injected. The incumbent strategy is said to be an ESS, if for each mutant strategy, there exists a positive invasion barrier such that if the population share of mutants falls below this barrier, the incumbents earns a higher expected payoff than the mutants. Consequently incumbents have no incentive to change their strategy, and the mutants have an incentive to return to the incumbent strategy.

To use game theory methods in real world problems, the exact values of the payoffs should be known. However, since the players usually encounter uncertainties and their information is incomplete, the crisp values of the payoffs are usually unknown. Hence the need arises as to study the games with uncertain payoffs. A way to deal with uncertainties associated with payoffs is to use the concept of fuzzy sets. Initially, fuzzy sets were introduced by Butnariu \cite{Butnariu1978,Butnariu1979} in non-cooperative game theory to model fuzzy games. Many studies on fuzzy game theory have been reported in the literature \cite{Vijay2005,Garagic2003,Li2010,Chakeri2013,Song1999,Chen2006,Bector2005}.
The readers can refer to \cite{Larbani2009} for the survey on fuzzy game theory. However, to the best of our knowledge, there isn't any study reported dealing with uncertainties in EGT. Particularly, in this paper, we propose the concept of Fuzzy ESS in symmetric two player games with fuzzy payoffs. In this regards, instead of answering whether a strategy is an ESS or not, each strategy is assigned a membership that describes the degree to which it possess the characteristics of ESS. The idea of fuzzy ESS captures the nature of ESS in a fuzzy payoff game,
and can be considered as a refinement of fuzzy Nash equilibrium proposed in \cite{Chakeri2013,Chakeri2010,Chakeri2008,Sharifian2010,Chakeri20081}. Specifically, we show that the fuzzy set of ESS is a subset of the fuzzy set of Nash equilibrium in symmetric two player games.

The remainder of this paper is organized as follows. Section \ref{s2} reviews the related concepts in EGT. The proposed approach to evaluate ESS in games with fuzzy payoffs is introduced in this section. It also contains the relation between the proposed fuzzy ESS and the fuzzy Nash equilibrium. In section \ref{s3}, we illustrate the proposed method by simple examples with small number of strategies. In section \ref{s4}, concluding remarks are given and main findings of the paper are highlighted.

\section{Evolutionary Stable Strategies: A Fuzzy Generalization Scheme}\label{s2}

%\subsection{Symmetric Two-person Game}
A normal game consists of a set of players, their strategies, and the payoffs available for all
combinations of players¡¯ strategies. A strategic game $G$ can be defined as
\begin{equation}
G_1 = (N, (S_i), (\$_i)),
\end{equation}
where $N=\{1,\ldots,n\}$ is the set of $n$ players, $S_i$ denotes the set of strategies available to the $i$th player and $\$_i: \prod_{j\in N}S_j \rightarrow \mathbb{R}$ is the payoff function of the $i$th player ($\prod_{j\in N}S_j$ is the set of all possible combination of strategies). In economics, the payoffs are usually firms' profits or consumers' utilities,
while in biology, payoffs usually represent individual fitness.

For a symmetric two-player game, i.e. $N=\{1,2\}$, the two players have the same strategy set $S=\{s_1,s_2,\ldots,s_{|S|}\}$, namely $S_1=S_2=S$,
where $|S|$ denotes the cardinality of $S$.
For any $i,j\in \{1,2,\ldots,|S|\}$,
the payoff for the player 1 playing $s_i$ facing the oponent playing $s_j$ is the same as
the payoff for the player 2 playing $s_i$ facing the oponent playing $s_j$, i.e.  $\$_1(s_i,s_j)=\$_2(s_j,s_i)$.
Hence the payoffs of the both players can be described using the same function $\$$.
If the player 1 plays $s_i$ and the player 2 plays $s_j$, the payoffs are $\$(s_i,s_j)$ to Player 1 and $\$(s_j,s_i)$ to Player 2. Thus a symmetric two-player game $G$ can be defined by
\begin{equation}
G = (S, \$).
\end{equation}
where $\$$ can be represented by a $|S| \times |S|$ payoff matrix.

For the simplicity of the notation, in the remainder of this paper, the index of a strategy is used to represent the strategy itself. For instance, $\$(i,j)$ represents $\$(s_i,s_j)$.

\subsection{Evolutionarily Stable Strategy}
In this subsection, we review briefly the main concepts and definitions in EGT.

\textbf{Definition 1} \cite{Weibull1995}: For a symmetric two-player game $G$, a strategy $s_i\in S$ is an ESS if
for every $j\neq i$, there exists $\overline{\varepsilon}_{ij}\in(0,1)$ such that
\begin{align}\label{eq-ess}
(1-\varepsilon)\cdot\$(i,i)+\varepsilon\cdot\$(i,j)
>(1-\varepsilon)\cdot\$(j,i)+\varepsilon\cdot\$(j,j)
\end{align}
holds when $\varepsilon\in(0,\overline{\varepsilon}_{ij})$.

Suppose that a population of individuals that plays strategy $s_i$,
is invaded by a group of mutants that play the alternative strategy $s_j$.
Equation \ref{eq-ess} requires that regardless of the choice of $s_j$, an incumbent's expected payoff (left hand side of the inequality)
from a random match in the post-entry population exceeds that of a mutant (right hand side of the inequality) so long as
the size of the invading group $\varepsilon$ is smaller than the invasion barrier $\overline{\varepsilon}_{ij}$.
A lower invasion barrier means that the population is more vulnerable to invasion,
while a higher invasion barrier means that the population is more stable.

For instance, consider the Prisoner's Dilemma game that is a well-known symmetric $2\times 2$ game,
representing the conflict between individual rationality and social optimality.
For this game,
$S=\{C,D\}$, where $C$ denotes ``Cooperate'' and $D$ denotes ``Defect''.
The payoffs satisfy $\$(D,C)>\$(C,C)>\$(D,D)>\$(C,D)$. According to definition 1, one can show that the strategy $D$ is an ESS. Hence, a player tends to defect rather than to cooperate.

\subsection{Fuzzy Evolutionarily Stable Strategy}
In this subsection, we generalize the ESS in games with fuzzy payoffs. In particular, a symmetric two-player game with fuzzy payoffs is similar to crips one
except that the payoffs are fuzzy numbers, i.e.
\begin{equation}
G = (S, \tilde{\$}).
\end{equation}
where $\tilde{\$}$ shows the set of fuzzy payoffs.

When the payoffs are not crisp values, the expected payoffs for the players have uncertainties. In this regard, the expected payoff of an incumbent playing strategy $i$ against a mutant playing strategy $j$ is a fuzzy number
%The fuzzy set of ESS for a fuzzy game $G$ can be obtained by the following steps:
%\\\\
%\noindent\emph{Step 1}.
\begin{align}\label{eq-epi}
\tilde{E}_I^{ij}(\varepsilon) = (1-\varepsilon)\cdot \tilde{\$}(i,i) + \varepsilon\cdot\tilde{\$}(i,j),
\end{align}

Similarly, the expected payoff for the mutant is
\begin{align}\label{eq-epm}
\tilde{E}_M^{ij}(\varepsilon) = (1-\varepsilon)\cdot \tilde{\$}(j,i) + \varepsilon\cdot\tilde{\$}(j,j).
\end{align}

Now, to compute the fuzzy expected values for the players, the multiplication and addition can be done using the extension principle \cite{Zimmermann1996}:
\begin{align}
& \mu_{\lambda \cdot A}(x) =  \mu_A(\frac{x}{\lambda}),\\
& \mu_{A + B}(z) =  \sup_{x+y=z}\min(\mu_A(x),\mu_B(y)),
\end{align}
where $A,B$ are fuzzy numbers, and $\lambda\in\mathbb{R}\setminus\{0\}$.

%\noindent\emph{Step 2}.
However, the question is how to compare fuzzy expexted payoffs. There are several different methods for comparing possibilistic distributions in the context of fuzzy set theory. In  this paper, in order to compare fuzzy numbers $\tilde{E}_I^{ij}(\varepsilon)$ and $\tilde{E}_M^{ij}(\varepsilon)$, we used the satisfaction function \cite{Lee2001}. The satisfaction function is the truth value for comparison between fuzzy values. In particular, the satisfaction function for the comparison between fuzzy values $A$ and $B$ is defined as
\begin{align}
 SF(A>B)=\frac{\int_{-\infty}^{\infty}\int_{y}^{\infty}\mu_A(x)\odot\mu_B(y) \mathrm{d}x \mathrm{d}y}{\int_{-\infty}^{\infty}\int_{-\infty}^{\infty}\mu_A(x)\odot\mu_B(y) \mathrm{d}x \mathrm{d}y},\\
 SF(A<B)=\frac{\int_{-\infty}^{\infty}\int_{-\infty}^{y}\mu_A(x)\odot\mu_B(y) \mathrm{d}x \mathrm{d}y}{\int_{-\infty}^{\infty}\int_{-\infty}^{\infty}\mu_A(x)\odot\mu_B(y) \mathrm{d}x \mathrm{d}y},
\end{align}
where $\odot$ is a T-norm operator that can be the min operator, multiplication, and etc. In this paper, we use multiplication operator as the T-norm to simplify the calculations.

$SF(A>B)$ represents the possibility that $A$ is larger than $B$,
while $SF(A<B)$ represents the possibility that $A$ is smaller than $B$. Clearly, $SF(A>B)+SF(A<B)=1$, and $SF(A>A)=0.5$.

Although some restrictions on $A,B$ are necessary \cite{Lee2001},
most types of fuzzy sets in different applications, such as triangular and trapezoidal, satisfy the imposed restrictions.

Now, by using the satisfaction function, the following relation represents the degree to which Equation \ref{eq-ess} is satisfied
\begin{equation}\label{eq-muij}
\mu_{ij}(\bar{\varepsilon})=\inf_{\varepsilon\in(0,\bar{\varepsilon})}SF(\tilde{E}_I^{ij}(\varepsilon)>\tilde{E}_M^{ij}(\varepsilon)).
\end{equation}

In addition, by applying Bellman and Zadeh's fuzzy decision rule \cite{Bellman1970}, we define the degree of resistibility of $s_i$ against $s_j$ as the supremum of the intersection of $\mu_{ij}$ and the invasion barrier $\bar{\varepsilon}$, i.e.
\begin{equation}\label{eq-rij}
r_{ij} = \sup_{\bar{\varepsilon}} \min(\mu_{ij}(\bar{\varepsilon}),\bar{\varepsilon} ).
\end{equation}

%\noindent\emph{Step 3}.
Finally, the fuzzy ESS set $\tilde{S}$ can be defined as
\begin{equation}\label{eq-mus}
\mu_{\tilde{S}}(i) =  \min_{j\neq i} r_{ij}.
\end{equation}
The membership function value $\mu_{\tilde{S}}(i)$ describes the degree to which the strategy $s_i$ possesses the characteristic of ESS. The strategies can be ranked according to their membership. Strategies with a higher membership are evolutionarily stronger than those with a lower membership.

\subsection{Relation between Fuzzy ESS and Fuzzy Nash Equilibrium}
In this subsection, the relation between the proposed generalization of the ESS and fuzzy Nash equilibrium \cite{Chakeri2013} is studied. In this regard, we review the Nash equilibrium concept and its fuzzy generalization, as in the following.

The concept of Nash equilibrium \cite{Fudenberg1991} is a cornerstone of non-cooperative game theory.
A strategy combination is called a Nash equilibrium (NE), if no player can do better by unilaterally changing his strategy. In other words, every player adopts his best response to the strategies of the other players.

\textbf{Definition 2}: For a strategic game $G = (N, (S_i), (\$_i))$, a strategy combination $(s_1^*,s_2^*,\ldots,s_n^*)\in \prod_{j\in N}S_j$ is a NE if
for every $i\in N$ and every $s_i\neq s_i^*\in S_i$,
\begin{equation}
\$_i(s_1^*\ldots s_i^* \ldots s_n^*)\geqslant \$_i(s_1^*\ldots s_{i-1}^*,s_i,s_{i+1}^*\ldots s_n^*)
\end{equation}.

For instance, $(D,D)$ is the unique NE in prisoner's dilemma game,
and is also called a symmetric NE because of the symmetry of the strategy pair.

In \cite{Chakeri2013}, the authors proposed an interesting approach on finding degree of being NE of each strategy combination in fuzzy payoff games.
Specifically, each strategy combination $(s_1,s_2,\ldots,s_n)$ has a possibility of being NE with the degree
\begin{align}\label{eq-fne}
\mu_{Nash}(s_1,\ldots,s_n)
= & \min_{\forall i\in N} \bigg( \min_{\forall s_i^{'}\neq s_i\in S_i} \Big(SF(\tilde{\$}_i(s_1\ldots s_i \ldots s_n)\nonumber\\
 & > \tilde{\$}_i(s_1\ldots s_{i-1},s_i^{'},s_{i+1}\ldots x_n)\Big)\bigg).
\end{align}

Although ESS and NE were proposed in quite different settings,
there exists a well established relation between ESS and symmetric NE in symmetric two-player games. That is, the symmetric NE set contains the ESS set \cite{Hofbauer1998}.
In this paper, we show that there is a similar relation between fuzzy ESS and fuzzy symmetric NE, as follows.

\textbf{Theorem 1}: In a symmetric two-player fuzzy game $G$, the fuzzy symmetric NE set contains the fuzzy ESS set, i.e. for every $i\in \{1,2,\ldots,|S|\}$,
\begin{equation}
\mu_{Nash}(s_i,s_i)\geqslant\mu_{\tilde{S}}(s_i).
\end{equation}

\textit{Proof}: According to Equation.~(\ref{eq-fne}),
\begin{align}
\mu_{Nash}(s_i,s_i)= & \min\bigg(  \min_{j\neq i} SF\Big(\tilde{\$}_1(s_i,s_i)>\tilde{\$}_1(s_j,s_i)\Big),\nonumber\\
& \min_{j\neq i} SF\Big(\tilde{\$}_2(s_i,s_i)>\tilde{\$}_2(s_i,s_j)\Big)\bigg)\nonumber  \\
= & \min\bigg(  \min_{j\neq i} SF\Big(\tilde{\$}(s_i,s_i)>\tilde{\$}(s_j,s_i)\Big), \nonumber\\
& \min_{j\neq i} SF\Big(\tilde{\$}(s_i,s_i)>\tilde{\$}(s_j,s_i)\Big)\bigg)\nonumber \\
= & \min_{j\neq i} SF\Big(\tilde{\$}(s_i,s_i)>\tilde{\$}(s_j,s_i)\Big).
\end{align}

According to Equations~\ref{eq-muij},\ref{eq-rij} and \ref{eq-mus},
\begin{align}
\mu_{\tilde{S}}(s_i)
= &\min_{j\neq i}\sup_{\bar{\varepsilon}}\min(\inf_{\varepsilon\in(0,\bar{\varepsilon})}SF(\tilde{E}_I^{ij}(\varepsilon)>\tilde{E}_M^{ij}(\varepsilon)),\bar{\varepsilon})\nonumber\\
\leqslant &  \min_{j\neq i}\sup_{\bar{\varepsilon}}\inf_{\varepsilon\in(0,\bar{\varepsilon})}SF(\tilde{E}_I^{ij}(\varepsilon)>\tilde{E}_M^{ij}(\varepsilon))\nonumber\\
= &  \min_{j\neq i}SF(\tilde{E}_I^{ij}(0)>\tilde{E}_M^{ij}(0))\nonumber\\
= &  \min_{j\neq i}SF(\tilde{\$}(s_i,s_i)>\tilde{\$}(s_j,s_i)).
\end{align}
Thus
\begin{equation}
\mu_{Nash}(s_i,s_i)\geqslant\mu_{\tilde{S}}(s_i).
\end{equation}
$\blacksquare$

As a result, the proposed fuzzy set of ESS can be considered as a refinement of the concept of fuzzy NE in \cite{Chakeri2013}.

\section{Numerical Results}\label{s3}

\begin{figure}
\centering
\includegraphics[width=2.5in]{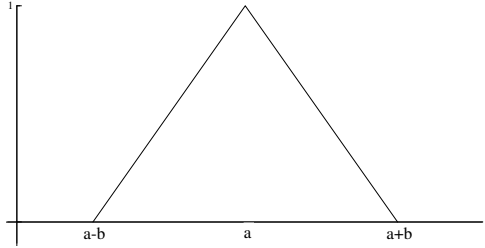}
\caption{The membership function of a symmetric triangular fuzzy number $T(a,b)$}
\label{fig:stfn}
\end{figure}

In this section, we present two examples to illustrate the proposed approach. Specifically, in \cite{Chakeri2013}, a two-player game with fuzzy payoffs were used to evaluate the fuzzy Nash equilibrium. The row player's payoffs of that game are collected in Table \ref{te1}, and is considered as the first example of a symmetric two-player game.
The column player's payoffs are collected in Table \ref{te2}, and is considered as the second  example.
For instance, for the firsr exaple, the payoff of a player when he plays $s_1$ and his oponent plays $s_2$ is $\tilde{\$}(1,2)=T(6,1)$.
$T(a,b)$ denotes a symmetric triangular fuzzy number with the center on $a$ and boundaries on $a\pm b$. Fig.~\ref{fig:stfn} shows the membership function of $T(a,b)$.
Multiplication and addition involving symmetric triangular fuzzy numbers using the extension principle are considerably simplified.
Let $T(a,b)$ and $T(a',b')$ be two symmetric triangular fuzzy numbers, and $c$ be a positive real number. Then
\begin{align}
& c\cdot T(a,b)=T(c a,c b),\label{eq-rule1}\\
& T(a,b)+T(a',b')=T(a+a',b+b').\label{eq-rule2}
\end{align}

Hence the expected payoffs $\tilde{E}_I^{ij}(\varepsilon)$ and $\tilde{E}_M^{ij}(\varepsilon)$ in Eq. (\ref{eq-epi}) and (\ref{eq-epm}) are also symmetric triangular fuzzy numbers.

\begin{table}[htb]
\caption{Payoff matrix of the first example} \label{te1}
\centering
\renewcommand{\arraystretch}{1.3}
\begin{tabular}{cccc}
\hline
$\tilde{\$}(i,j)$ & $s_1$ & $s_2$ & $s_3$ \\
\hline
$s_1$ & $T(5,1)$ & $T(6,1)$ & $T(5,2)$ \\
$s_2$ & $T(3,1)$ & $T(3,2)$ & $T(3,1)$ \\
$s_3$ & $T(4,1)$ & $T(5,2)$ & $T(7,1)$ \\
\hline
\end{tabular}
\end{table}

\begin{table}[htb]
\caption{Payoff matrix of the second example} \label{te2}
\centering
\renewcommand{\arraystretch}{1.3}
\begin{tabular}{cccc}
\hline
$\tilde{\$}(i,j)$ & $s_1$ & $s_2$ & $s_3$ \\
\hline
$s_1$ & $T(3,2)$ & $T(1,1)$ & $T(4,2)$ \\
$s_2$ & $T(3,1)$ & $T(4,1)$ & $T(3,2)$ \\
$s_3$ & $T(3,1.5)$ & $T(3,2)$ & $T(6,2)$ \\
\hline
\end{tabular}
\end{table}

In the first example, the degree of resistibility of the strategies against alternative strategies are
\begin{align}
& r_{12}=1, & r_{13}=0.397, \nonumber\\
& r_{21}=0, & r_{23}=0.034, \nonumber\\
& r_{31}=0.603, & r_{32}=0.966.
\end{align}

As a result, the fuzzy ESS set is as in equation \ref{eq-numeric}. Hence, according to their membership, the strategies can be ranked as $s_3>s_1>s_2$.
\begin{equation}\label{eq-numeric}
\tilde{S}=\frac{0.397}{s_1} + \frac{0}{s_2} + \frac{0.603}{s_3}.
\end{equation}

\begin{figure}
\centering
\includegraphics[width=2.5in]{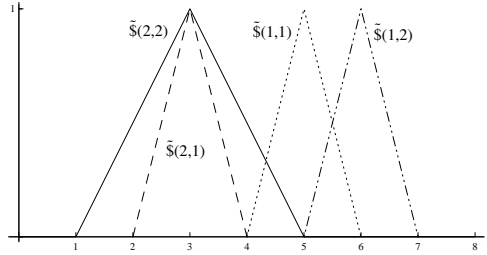}
\caption{Membership functions of $\tilde{\$}(1,1), \tilde{\$}(2,1), \tilde{\$}(1,2)$ and $\tilde{\$}(2,2)$.}
\label{fig:compare}
\end{figure}

Consider the situation in which a population playing strategy $s_1$ is invaded by a group of mutants playing $s_2$.
As can be seen in Fig.~\ref{fig:compare},
the left boundary of $\tilde{\$}(1,1)$ is not smaller than the right boundary of $\tilde{\$}(2,1)$,
and the left boundary of $\tilde{\$}(1,2)$ is not smaller than the right boundary of $\tilde{\$}(2,2)$.
According to Equations ~\ref{eq-rule1} and \ref{eq-rule2}, the left boundary of $(1-\varepsilon)\cdot\tilde{\$}(1,1)+\varepsilon\cdot\tilde{\$}(1,2)$ is not smaller than
the right boundary of $(1-\varepsilon)\cdot\tilde{\$}(2,1)+\varepsilon\cdot\tilde{\$}(2,2)$.
That is,
\begin{equation}
SF(\tilde{E}_I^{12}(\varepsilon)>\tilde{E}_M^{12}(\varepsilon))=1.
\end{equation}
Similarly,
\begin{equation}
SF(\tilde{E}_I^{21}(\varepsilon)>\tilde{E}_M^{21}(\varepsilon))=0.
\end{equation}
This explains the values of $r_{12}$ and $r_{21}$.

In the second example, the degree of resistibility of the strategies against alternative strategies are
\begin{align}
& r_{12}=0.222, & r_{13}=0.295, \nonumber\\
& r_{21}=0.778, & r_{23}=0.349, \nonumber\\
& r_{31}=0.705, & r_{32}=0.651.
\end{align}

As a result, the fuzzy ESS set is
\begin{equation}
\tilde{S}=\frac{0.222}{s_1} + \frac{0.349}{s_2} + \frac{0.651}{s_3}.
\end{equation}

According to their membership degree, the strategies can be ranked as $s_3 > s_2 > s_1$.

Table \ref{tc} summarizes the fuzzy ESS sets and the fuzzy symmetric NE sets of the two examples. For instance, $\mu_{Nash}(s_1,s_1)=0.958$, $\mu_{Nash}(s_2,s_2)=0$, $\mu_{Nash}(s_3,s_3)=0.989$ in the first example.
It can be seen that a strategy with a high degree of being an ESS also has a high degree of being a symmetric NE,
while a strategy with a high degree of being a symmetric NE doesn't necessarily have a high degree of being an ESS.
This comparison result is congruous with the relation between fuzzy ESS and fuzzy symmetric NE.

\begin{table}[htb]
\caption{Comparison between fuzzy ESS and fuzzy NE} \label{tc}
\centering
\renewcommand{\arraystretch}{1.3}
\begin{tabular}{cccc}
\hline
&  $s_1$ & $s_2$ & $s_3$ \\
\hline
Fuzzy ESS of the first game  & $0.397$ & $0$ & $0.603$  \\
Fuzzy NE of the first game   & $0.958$ & $0$ & $0.989$ \\
Fuzzy ESS of the second game & $0.222$ & $0.349$ & $0.651$ \\
Fuzzy NE of the second game  & $0.5$ & $0.854$ & $0.958$ \\
\hline
\end{tabular}
\end{table}

\section{Discussion and Conclusion}\label{s4}

A key concept in evolutionary game theory is that of an evolutionarily stable strategy (ESS).
It's defined by comparison between an incumbent's expected payoff and a mutant's expected payoff.
When the payoffs are represented by fuzzy numbers, the comparison results become vague,
so there is no way to determine whether a strategy is an ESS.
Although many works on fuzzy games have been done,
there has been no generalization of ESS into a fuzzy one.
Inspired by the recent work on fuzzy Nash equilibrium \cite{Chakeri2013},
this paper proposes a new approach to evaluate ESS in games with fuzzy payoffs.

In this method, the possibility that an incumbent's expected payoff is larger than a mutant's, is a function of $\bar{\varepsilon}$ (the invasion barrier).
This function $\mu_{ij}(\bar{\varepsilon})$ incorporates the satisfaction function, which is also employed to compare two fuzzy payoffs in the fuzzification process of NE.
Besides $\mu_{ij}(\bar{\varepsilon})$, $\bar{\varepsilon}$ is also an indicator of population stability,
hence the degree of resistibility of one strategy against another, denoted as $r_{ij}$, is defined by using fuzzy decision rule on $\mu_{ij}(\bar{\varepsilon})$ and $\bar{\varepsilon}$.
The membership function $\mu_{\tilde{S}}(i)$ is defined as the minimum of $r_{ij}$.

Since we used a similar point of view to fuzzy ESS as in \cite{Chakeri2013} to the concept
of fuzzy NE, the concept of fuzzy ESS has close relation to that of fuzzy NE.
Just like crisp ESS can be considered as a refinement of crisp NE \cite{Weibull1995},
fuzzy ESS can be considered as a refinement of fuzzy NE.
It is worthwhile to note that both fuzzy NE and fuzzy ESS assign membership to pure strategies. The formalizations of fuzzy NE and fuzzy ESS concerning mixed strategies are considerably complicated due to the continuity of mixed strategies.

Also, we address the relation between crisp ESS and fuzzy ESS. Specifically, what if the proposed approach is applied to crisp payoff games?
When the payoffs are crisp numbers,
the expected payoffs of incumbents and mutants are the following crisp numbers:
\begin{align}
& E_I^{ij}(\varepsilon) = (1-\varepsilon)\cdot \$(i,i) + \varepsilon\cdot\$(i,j),\\
& E_M^{ij}(\varepsilon) = (1-\varepsilon)\cdot \$(j,i) + \varepsilon\cdot\$(j,j).
\end{align}
In this regard, the satisfaction function for the comparison of crisp values $k$ and $l$ is defined as \cite{Lee2001}
\begin{align}
 SF(k>l)=\begin{cases}
1,    & k>l  \\
0.5,    & k=l  \\
0,   & k<l
\end{cases},\\
 SF(k<l)=\begin{cases}
1,    & k<l  \\
0.5,    & k=l  \\
0,   & k>l
\end{cases}.
\end{align}

Hence, it is straightforward to verify that
$r_{ij}$ is equal to the supremum of $\overline{\varepsilon}_{ij}$ if $\overline{\varepsilon}_{ij}$ exists,
and is equal to $0.5$ if $E_I^{ij}(\varepsilon)=E_M^{ij}(\varepsilon)$ for every $\varepsilon\in(0,1)$,
and is equal to $0$ in other cases.
As an example, consider the Stag Hunt game that describes a conflict between safety and social cooperation.
Two hunters go out on a hunt. Each can choose to hunt a stag or a hare (represented by strategies $G$ and $H$ respectively).
A hunter who hunts hare obtains a payoff of $h>0$.
A hunter who hunts stag obtains a payoff of $g>h$ if the other hunter also hunts stag,
and obtains a payoff of $0$ otherwise.
Formally, $\$(H,H)=\$(H,G)=h,\$(G,G)=g,$ and $(G,H)=0$.
According to the definition of crisp ESS, both $G$ and $H$ are ESS.
By contrast, the fuzzy ESS set for Stag Hunt is
\begin{align}
& \mu_{\tilde{S}}(H)=\frac{h}{g},\\
& \mu_{\tilde{S}}(G)=1-\frac{h}{g},
\end{align}
which gives a ranking of these two strategies.
When $h/g$ is small and $\mu_{\tilde{S}}(G)>\mu_{\tilde{S}}(H)$, the hunters tend to cooperate to hunt stag.
When $h/g$ becomes larger and $\mu_{\tilde{S}}(H)>\mu_{\tilde{S}}(G)$, the hunters tend to hunt hare to avoid the risk of being betrayed.
In general, the membership function of a fuzzy ESS set for strategy $s_i$ is reduced to the minimum of the invasion barriers of $s_i$ against all $s_j (j\neq i)$.
In this way the idea of fuzzy ESS can be considered as a natural extension of the crisp ESS.
\\
\\
\noindent \textbf{Acknowledgement}
This work is supported by the National Natural Science Foundation of China (Grant Nos. 61502179, 61472452) and
the Natural Science Foundation of Guangdong Province of China (Grant No. 2014A030310265).


\begin{thebibliography}{1}
\bibitem{Basar1998} T. Basar, G.J. Olsder. \emph{Dynamic Noncooperative Game Theory, 2nd Edition}.
Society for Industrial and Applied Mathematics, 1998.
\bibitem{Fudenberg1991} D. Fudenberg, J. Tirole. \emph{Game Theory}. MIT Press, 1991.
\bibitem{Weibull1995} J.W. Weibull. \emph{Evolutionary game theory}. MIT Press, 1997.
\bibitem{Hofbauer1998} J. Hofbauer, K. Sigmund. \emph{Evolutionary Games and Population Dynamics}. Cambridge University Press, 1998.
\bibitem{Maynard1973} J.M. Smith, G.R. Price. The logic of animal conflict.
Nature, 246: 15-18, 1973.
\bibitem{Butnariu1978} D. Butnariu. Fuzzy games: a description of the concept.
Fuzzy Sets and Systems, 1(3): 181-192, 1978.
\bibitem{Butnariu1979} D. Butnariu. Solution concepts for n-person fuzzy games.
Advances in Fuzzy Set Theory and Applications, 339-354, 1979.
\bibitem{Vijay2005} V. Vijay, S. Chandra, C.R. Bector. Matrix games with fuzzy goals and fuzzy payoffs.
Omega, 33(5): 425-429, 2005.
\bibitem{Garagic2003} D. Garagic, J.B. Cruz Jr. An approach to fuzzy noncooperative Nash games.
Journal of Optimization Theory and Applications, 118(3): 475-491, 2003.
\bibitem{Li2010} D.F. Li. Mathematical-programming approach to matrix games with payoffs represented by Atanassov's interval-valued intuitionistic fuzzy sets. IEEE Transactions on Fuzzy Systems, 18(6): 1112-1128, 2010.
\bibitem{Chakeri2013} A. Chakeri, F. Sheikholeslam. Fuzzy Nash equilibriums in crisp and fuzzy games.
IEEE Transactions on Fuzzy Systems, 21(1): 171-176, 2013.
\bibitem{Song1999} Q. Song, A. Kandel. A fuzzy approach to strategic games. IEEE Transactions on Fuzzy Systems, 7(6): 634-642, 1999.
\bibitem{Chen2006} Y.W. Chen, M. Larbani. Two-person zero-sum game approach for fuzzy multiple attribute decision making problems. Fuzzy Sets and Systems, 157(1): 34-51, 2006.
\bibitem{Bector2005} C.R. Bector, S. Chandra. \emph{Fuzzy Mathematical Programming and Fuzzy Matrix Games.}
Vol. 169. Springer, 2005.
\bibitem{Larbani2009} M. Larbani. Non cooperative fuzzy games in normal form: A survey.
Fuzzy Sets and Systems, 160(22): 3184-3210, 2009.
\bibitem{Chakeri2010} A. Chakeri, N. Sadati, S. Sharifian. Fuzzy Nash equilibrium in fuzzy games using ranking fuzzy numbers. Fuzzy Systems (FUZZ), 2010 IEEE International Conference on, pp. 1-5.
\bibitem{Chakeri2008} A. Chakeri, A.N. Dariani, C. Lucas. How can fuzzy logic determine game equilibriums better?
Intelligent Systems, 2008. IS '08. 4th International IEEE Conference  (Volume:1 ), pp. 2-51-2-56.
\bibitem{Sharifian2010} S. Sharifian, A. Chakeri, F. Sheikholeslam. Linguisitc representation of Nash equilibriums in fuzzy games. Fuzzy Information Processing Society (NAFIPS), 2010 Annual Meeting of the North American, pp. 1-6.
\bibitem{Chakeri20081} A. Chakeri, J. Habibi, Y. Heshmat. Fuzzy type-2 Nash equilibrium.
Computational Intelligence for Modelling Control Automation, 2008 International Conference on, pp. 398-402.
\bibitem{Zimmermann1996} H.-J. Zimmermann. \emph{Fuzzy set theory¡ªand its applications, 4th Edition}.
Kluwer Academic Publishers, 2001.
\bibitem{Lee2001} J.H. Lee, H. Lee-Kwang. Comparison of fuzzy values on a continuous domain.
Fuzzy Sets and Systems, 118(3): 419-428, 2001.
\bibitem{Bellman1970} R.E. Bellman, L.A. Zadeh. Decision-making in a fuzzy environment.
Management Science, 17(4): B-141-B-164, 1970.
\end{thebibliography}
\end{document}